# Early Career Perspectives For the NASA SMD Bridge Program

**Leads:**
Dr. Jenna M. Cann (NASA Goddard Space Flight Center/ORAU)
Dr. Arturo O. Martinez (Bay Area Environmental Research Institute/NASA Ames)

**Members:**
Amethyst Barnes (Florida Institute of Technology)
Sara Doan (George Mason University)
Dr. Feyi Ilesanmi (Gordonville Management Consulting)
Dr. Margaret Lazzarini (Caltech, California State University Los Angeles)
Dr. Teresa Monsue (NASA Goddard Space Flight Center/Catholic University of America)
Lt. Col. Carlos Pinedo, Ph.D. (US Air Force Test Pilot School)
Dra. Nicole Cabrera Salazar (Movement Consulting)
Dr. Amy Steele (McGill University)

**Abstract:**
In line with the Astro2020 Decadal Report "State of the Profession" findings and the NASA core value of Inclusion, the NASA Science Mission Directorate (SMD) Bridge Program was created to provide financial and programmatic support to efforts that work to increase the representation and inclusion of students from under-represented minorities in the STEM fields. To ensure an effective program, particularly for those who are often left out of these conversations, the NASA SMD Bridge Program Workshop was developed as a way to gather feedback from a diverse group of people about their unique needs and interests. The Early Career Perspectives Working Group was tasked with examining the current state of bridge programs, academia in general, and its effect on students and early career professionals. The working group, comprised of 10 early career and student members, analyzed the discussions and responses from workshop breakout sessions and two surveys, as well as their own experiences, to develop specific recommendations and metrics for implementing a successful and supportive bridge program. In this white paper, we will discuss the key themes that arose through our work, and highlight select recommendations for the NASA SMD Bridge Program to best support students and early career professionals.

**Introduction:**
The *Early Career Perspectives Working Group* (ECP WG) was formed with the aim of determining the primary challenges, goals, and concerns of the early career population and relate them back to the goals, ideals, and best practices for the *NASA SMD Bridge Program*. In order to ensure a robust exploration of these topics, *ECP WG* members were recruited from across scientific disciplines and career paths. This white paper will address themes that are very similar to those of previous bridge programs, specifically the Fisk-Vanderbilt and Cal-Bridge programs, and will highlight places where future bridge programs can improve upon existing models.

The findings presented here arise from three main sources: intra-working group conversations, the Early Career Perspectives breakout sessions facilitated by working group chairs and members at the October 2022 SMD Bridge Program Workshop, and two anonymous surveys, one geared towards current students and one towards recent alumni working in various industries. These findings are organized into four main themes: mentorship, career development, formal support systems, and the need for an inclusive community. In the final section of this report, we highlight specific recommendations for bridge programs to best support their students throughout their time in the program and in their future careers.

## Mentorship:

A growth-mindset mentorship program needs to be an integral part of the program structure, designed as an asset-based, rather than a deficit-based, model that can foster a healthy academic environment [Dweck 2006]. Students should have access to a variety of mentors that guide them through the course of their degree(s), advocate for them in cases of concerns, and ensure they are well-prepared to meet their career goals. The professional mentors that are assigned to the students should be aligned with their ideal career path for the program to be most effective and useful for their career development, as mentoring programs in general have been proven to improve student performance [Cross & Vick 2001]. In particular, robust mentorship programs have been shown to increase retention, strengthen feelings of belonging, and prepare students for their chosen career path [Godwin et al., 2016]. We emphasize that students should be thoughtfully matched with mentors at the onset of the program, and they should be matched with more than a single mentor to ensure that they can receive advice from a variety of professionals and viewpoints.

Following a similar model from the Cal-Bridge program [Rudolph 2019, Rudolph et al. 2019], we recommend that a main mentor at the host institution look after the student's day-to-day development and can take on the role of a research and/or career advisor. A secondary mentor/advisor would look after the students' career development on a week-to-week basis and would be essential in providing more than one perspective on the student's career goals. There are also optional recommended mentors that could take on additional different specific roles. For example, a mentor with a similar background (e.g., ethnicity, gender, national origin, educational background, field of study, etc.) to the student could help ensure the student feels comfortable, included, and supported throughout their experience at the bridge program and offer advice regarding specific concerns that someone with a different background may not share [Trawick et al. 2021].

There also should be opportunities for peer and near-peer mentorship. This mentorship can take many forms, including opportunities for more senior students to mentor new students in groups or clusters, providing new students with community and helping new mentors support each other. While the ideal scheme will depend on the specifics of the institution, we suggest as an example that as the program progresses, prior and more senior student members could participate as mentors for incoming students. Aside from the social, mental, and professional benefits to the mentee, the opportunities offered to mentors encourages new students to meaningfully engage as a leader in their community, thus increasing their feelings of belonging and commitment to the program and the field [Pando 2022]. The Fisk-Vanderbilt shows effective peer mentoring in the "bridge buddy system", where incoming students are paired up with more established students. Discussions on this program highlighted participants' increased comfort with talking to their peers about difficult issues (e.g., harassment, department politics, imposter syndrome, etc.) that they were less comfortable discussing with a faculty or professional mentor. By providing peer and near-peer mentorship, just like other bridge programs are currently doing or have done in the past [Rudolph 2019, Rudolph et al. 2019], it can help keep student retention rates high among *NASA SMD Bridge Program* scholars [e.g., Hurtado et al. 2010].

We emphasize that mentorship training is a necessary part of an effective mentorship program. Mentorship is a skill that must be developed, however there is often little funding, time, or support set aside to empower new mentors. This is a significant concern, as a poor or inexperienced mentor could inadvertently cause harm to the mentee [Cabrera Salazar 2023]. To truly address these concerns, we recommend extended training that focuses not only on traditional mentorship, but also on the effects that systemic oppression and bias present in academic settings may have on the mentor-mentee relationship. It is also important to provide opportunities for peer support of mentors, and offer opportunities for mentors to discuss concerns specific to a given field or department, as well as provide advice and motivation for each other. Unless absolutely necessary, we emphasize that all mentors should discuss all their mentees' concerns with full anonymity in order to respect the privacy of each scholar. Finally, we also recommend separate mentorship-related programming for mentees as well (e.g., how to get the most out of your mentorship experience) to empower them in this relationship to ensure that their needs are being met, and to know where to bring concerns if they are not.

**Career Development:**

Students from underrepresented groups must be supported throughout their education and be able to gain the skills to thrive in a wide range of careers that meet their needs and goals. The ratio of academic research positions to PhDs is too low to reasonably accommodate all interested graduates. Additionally, there are multiple fulfilling career paths that STEM graduates can take, and there has been a strong shift in interest towards non-academic pathways in recent years. The career development opportunities and resources offered by many departments, however, have not kept up with the realities of the academic job market. To fully support students, it is important to ensure that their degree programs are helping prepare them for their ideal career pathway, including and especially non-academic careers.

This support can take many forms, including skills-based programming, networking with alumni in various jobs, and internship opportunities. In conversations and survey responses [Cann & Martinez 2022], students and early career scientists were asked to broadly state career preparedness and support from networks or mentors, whether it was in academia or the general workforce. They cited a lack of formal instruction to prepare for career paths other than as professors or research scientists. In particular, common concerns included a lack of experience or instruction in "real world" skills like coding, scientific writing, machine learning, and project management. The development of skills-based workshops or other resources, particularly with the input of professionals actively working in those fields, would ensure that students feel more confident and prepared to enter the workforce after graduation.

We recognize that the development of robust skill-building workshops can require significant effort and time commitment, and we emphasize the benefits of creating strong institutional networks, both with other local institutions to develop materials tailored to graduates in a specific local area or broadly relevant to relevant majors, either virtually or hybrid, with the help of professional societies, industry, or non-profit partners. We also emphasize the benefit of keeping in contact with alumni who can provide near-peer advice and insight into diverse non-academic jobs to current students. While the use of alumni networks is not uncommon in general, many institutions leave it up to the student to initiate the conversation or rely on faculty members to make the connection. This can result in researchers who already have a relatively strong network gaining even more resources than those who do not. To mitigate this, we recommend formalizing this process by hosting networking events each semester or year that are widely advertised and open to all students and alumni.

Internships and research opportunities can provide valuable insight for the student into the work culture and environment of a prospective post-graduation employer or career path. However, the process for obtaining an internship or research opportunity can often be opaque for those without existing social connections or job-search experience. In addition, unpaid or low-pay positions are inaccessible for many underrepresented students due to financial need. We recommend incorporating a widely publicized, financially-supported internship or research experience into the program, or providing separate financial support for a student if they choose to take an internship or research experience that would otherwise be unpaid or low-pay. This experience should be geared as much as possible to student interests to ensure that it will be a fulfilling experience that prepares them for their ultimate career goals.

Improving career development for students also requires dedicated effort and education of mentors in how best to support the student's unique professional development goals, as many faculty have either outdated or no experience in a non-academic position. Even the job search process itself can be very different between academic and non-academic jobs, and the resources and faculty expertise available to many students is often limited to what is relevant for academic positions. Specific concerns identified in survey responses [Cann & Martinez 2022] along this thread included interview preparation, resume and cover letter writing, and general job application advice. The development of internal resources and/or the use of external experts may be necessary to ensure that students are prepared and confident for their job search.

It is also important to consider the career development of students who take "non-traditional" career paths. Many scientists do not take the linear path from high school to college to graduate school (e.g., 2-year college to 4-year university, 4-year university to graduate school, industry to 2-year college/4-year university/graduate school), and these steps may happen at any age or career stage. It is vitally important to ensure that these students can fully take advantage of the opportunities provided through the *NASA SMD Bridge Program* to the extent that they wish (with very minimal age restrictions), and that the institutions in the *NASA SMD Bridge Program* respect and celebrate the unique expertise and skills that they bring.

Finally, it is important for the *NASA SMD Bridge Program* to proactively push back against the misguided belief that leaving academia, at any stage of the process, is a sign of "failure". The purpose of the Bridge Program should be to prepare students for a career that fulfills them, and it is important to remove any preconceived notions of what success may look like. While the resources above will help promote this idea, faculty, administrators, and the *NASA SMD Bridge Program* messaging should emphatically celebrate the success of a Bridge Program participant achieving their career goals, including and especially those outside of academia, regardless of what those goals may be (i.e., any degree level to industry/teaching/etc).

## Formal Support Systems:

Students should be provided with the holistic support needed to complete their degree, regardless of personal circumstances or their support system/resources coming into the program. As there are many forms of support systems, we focus on three major topics:

(1) financial support, (2) academic/career-related and personal counseling, and (3) academic, career-related, and personal skill building.

Many undergraduate students, especially those from underrepresented groups, find themselves struggling to afford higher education along with everyday costs of living (e.g., rent, food, etc.). Students from low socioeconomic backgrounds may work outside of school up to 20-30 hours per week [Rudolph 2019], which can prevent them from devoting enough time to their academic or career development. Thus, adequate financial support will enable students to more fully participate in their education without compromising their financial safety. Relevant support could include, but should not be limited to: tuition, living expenses, reimbursements for graduate admission exams (or exams of a similar nature), and travel funds to attend conferences to showcase research and/or find career advancement opportunities.

The negative mental health effects of graduate study have been well studied [e.g., Woolston 2019, Forrester 2020, Murguía Burton & Cao 2022], with findings showing notably higher rates of depression and anxiety symptoms in grad students than the general population. Furthermore, a 2019 survey of PhD students by Nature found that approximately 20% of respondents sought help but did not feel supported by the services at their institution, while 10% said that there were no resources available at their institution [Woolston 2019]. Access to counseling can be beneficial for students at any bridge program, whether it comes in the form of personal or academic/career counseling. Since academic rigor, personal hardships, and other life factors can become significant roadblocks in a student's academic career goals, access to a personal counselor, or trained mental health mentor, can provide another avenue to ensure a healthy academic environment for the individual and the program as a whole [JED 2021]. Academic and career counseling would assist students in transitioning into their next career stage (e.g., career counseling, grad school applications, postdoc applications).

Skill-building workshops are essential for a bridge program in order to support their students to succeed. These can be split into two main types of workshops that offer support: academic and career-related workshops (e.g., coding workshops), and personal/life development workshops (e.g., time management and growth mindset). These types of workshops not only build a student's career but can also serve as a support system since they provide resources and can create a positive, nurturing work environment. Specifically, workshops or seminars that move away from the deficit-based model can provide cultural awareness to mentors. We emphasize that the NASA SMD Bridge Program should offer funding to institutions to compensate facilitators for these workshops.

Furthermore, underrepresented students can encounter various circumstances (e.g., disability, parenthood, financial duress) that may affect their consistent participation in the bridge program. An ideal program should adapt to the student's personal needs, and provide support and resources if they need to pause participation in the program.

## Inclusive Community:

It is of the utmost importance that all students and early career scientists work in a safe, inclusive environment where they can succeed. The status quo of the field promotes an unhealthy, extremely competitive environment. This problem can be exacerbated in bridge programs where students may need to compete for limited spaces as well as countering biases within their department and the field. Further, advisors and mentors may suffer from implicit affinity bias and subconsciously prioritize students who are from similar backgrounds, indirectly imposing a "two-tiered" system between students enrolled in the bridge program and those enrolled directly in graduate school. It is therefore necessary to provide robust implicit bias and inclusive communication training to all who will be serving in a support or advisory role to students.

To counteract these negative influences, it is important to actively support students in forming their own community, including providing financial and administrative support for counterspaces and cohort-bonding activities [e.g., Ong et al. 2018]. It is necessary to empower students to navigate and mitigate the systemic issues, as well as actively counter the "deficit" mentality behind the structure of many current bridge programs [Dweck 2006]. Therefore, we recommend the messaging and programmatic structure of the NASA SMD Bridge Program to explicitly combat these systemic issues. Possible initiatives could include, but are not limited to, a workshop series dedicated to student empowerment (e.g., talks on community building, self-advocacy, etc.), regular cohort lunches or social events, affinity groups for students to talk with people of similar backgrounds, and opportunities to interact with alumni to learn both about their career and their personal journey. We recognize the time and financial commitment necessary to put on a workshop series and other initiatives is substantial, and therefore recommend that NASA, or a regional partnership, provides significant support in this endeavor, both financially and administratively.

To ensure an inclusive working environment, we emphasize the need for safe, effective, and supportive mechanisms to report instances of misconduct. There is often a great deal of stress and uncertainty involved when making a report, particularly for students and early career scientists, whose career trajectory depends significantly on their advisor's support. Any organization working with the NASA SMD Bridge Program should provide multiple channels to make a report, including anonymous and confidential options. Furthermore, there must be accountability to ensure that student concerns are taken seriously and their safety is prioritized. This can take many different forms depending on the nature of the offense, including, but not limited to, reassigning of mentor/mentee pairs, loss of student supervision privileges, recommendation for additional training, or dismissal from the program.

Finally, in order to provide services that serve underrepresented populations, there is a need to provide official documentation in languages other than English. The US Census Bureau has found that in 2019, about 22% of the US population above five years old spoke a language other than English in their household [US Census 2022]. Within this 22% of the population, an approximate 62% are Spanish speakers, which is not surprising since Hispanic people are the largest ethnic minority group within the United States [US Census 2021]. However, resources must not be limited to Spanish and should include other languages. According to the US Census Bureau [US Census 2022], the top 10 languages/dialects spoken other than English, with their respective percentage, are the following: Spanish (61.6%), Chinese (5.2%; including Mandarin and Cantonese), Tagalog (2.6%; including Filipino), Vietnamese (2.3%), Arabic (1.9%), French (1.7%; including Cajun), Korean (1.6%), Russian (1.4%), Haitian (1.4%), and German (1.3%). However, this neglects to include any of Native American population. Therefore, we recommend that official documentation be written in these languages along with Navajo (at the very least, being the top spoken Native American language), as these top 10 languages/dialects (with the exception of Navajo) comprises of 81% of the non-English spoken languages within the US.

**Recommendations in line with Guiding Principles:**
In line with the core tenants of the *NASA SMD Bridge Program*, the *ECP WG* makes the following recommendations to ensure that student participants in the Bridge Program are well-supported and able to thrive. We understand that there are inherent inequities in the scientific community and higher education system. We are trying to mitigate these inequities through these recommendations, even if we cannot solve them through this program alone.

1) Provide and require adequate mentorship and implicit bias training for all mentors and leadership involved with the Bridge Program. To ensure that training is effective, sessions, workshops, and check-ins should be offered by professionals at regular intervals to ensure that mentors are well-equipped to handle needs/concerns from a diverse student population.
2) Prioritize support to cohort-building experiences, student empowerment workshops, and counterspaces for students to form strong support networks to counteract the negativity and toxicity currently inherent in academic environments.
3) Provide support for diverse career pathways (e.g., career counseling, skill development, workshops, mentorship), including and especially those outside of academia (e.g., industry, teaching, policy, international development, governance, etc.) since there are many STEM graduates pursuing non-academic careers [Porter & White 2019]. Many mentors would likely not have experience in careers outside of academia, so this therefore may require the involvement of external groups.

Furthermore, the *ECP WG* suggests that many of the resource-heavy recommendations be handled centrally, either through NASA, or through regional collaborations between Bridge Program institutions, to ensure that access is not dependent on an individual institution's potentially limited resources.

## Conclusion:

The findings and recommendations above represent the compilation of work done by all *ECP WG* members at the request of the Bridge Program Workshop Organizing Committee. We emphasize that these themes and findings, while informed by the diverse experiences of working group members, should be taken as a starting point for discussions specifically geared to a given institution. In particular, we recommend consulting with various stakeholders (e.g., students, faculty, alumni, administrative resources, professional societies, and local communities) to develop a Bridge Program geared to the unique needs and goals of each institution. Furthermore, while many of these conversations have taken place recently during the Bridge Program Workshop and related discussions, they should continue throughout any program to ensure that it remains adaptable to the evolving needs of the student population. Centering early career and student voices, which are often overlooked, in this endeavor will help ensure an effective and supportive program that truly works for the people it is created for.